\documentclass[aps,prc,twocolumn,10pt, superscriptaddress, showpacs, floatfix]{revtex4-1}
\usepackage{amssymb,epsfig}
\usepackage{amsmath,bm}
\usepackage{mathtools}
\newcommand{\be}{\begin{equation}}
\newcommand{\ee}{\end{equation}}
\newcommand{\bea}{\begin{eqnarray}}
\newcommand{\eea}{\end{eqnarray}}
\newcommand{\mathscr}{\cal}

\begin{document}

\title{Temperature evolution of the nuclear shell structure and the dynamical nucleon effective mass}
\author{Herlik Wibowo}
\affiliation{Department of Physics, Western Michigan University, Kalamazoo, MI 49008, USA}
\author{Elena Litvinova}
\affiliation{Department of Physics, Western Michigan University, Kalamazoo, MI 49008, USA}
\affiliation{National Superconducting Cyclotron Laboratory, Michigan State University, East Lansing, MI 48824, USA}
\affiliation{GANIL, CEA/DRF-CNRS/IN2P3, F-14076 Caen, France}
\author{Yinu Zhang}
\affiliation{Department of Physics, Western Michigan University, Kalamazoo, MI 49008, USA}
\author{Paolo Finelli}
\affiliation{Dipartimento di Fisica e Astronomia,
Universit\'a degli Studi di Bologna and
INFN, Sezione di Bologna, Via Irnerio 46, I-40126 Bologna, Italy}


\date{\today}

\begin{abstract}
We study the fermionic Matsubara Green functions in medium-mass nuclei at finite temperature. The single-fermion Dyson equation with the dynamical kernel of the particle-vibration-coupling (PVC) origin is formulated and solved in the basis of Dirac spinors, which minimize the grand canonical potential with the meson-nucleon covariant energy density functional. The PVC correlations beyond mean field are taken into account in the leading approximation for the energy-dependent self-energy, and the full solution of the finite-temperature Dyson equation is obtained for the fermionic propagators. Within this approach, we investigate the fragmentation of the single-particle states and its evolution with temperature for the nuclear systems $^{56,68}$Ni and $^{56}$Fe relevant for the core-collapse supernova. The energy-dependent, or dynamical, nucleon effective mass is extracted from the PVC self-energy at various temperatures. 

\end{abstract}

\maketitle

\section{Introduction} 

The problem of predictive description of strongly-correlated many-body systems remains at the frontiers of science for decades.  
Although its solutions have been lately boosted by the progress in numerical computation, there is still a need of conceptual and formal advancements in the related areas of physics. While finding exact solutions is not possible in principle, novel ideas to approach these solutions without referring to the perturbation theory are actively discussed.

One of the systematic ways is offered by the equation of motion (EOM) framework. The EOM's can be straightforwardly generated for various quantum mechanical quantities, for instance, the correlation functions of field operators. The simplest correlation functions are of the propagator type, which are related to the spectral characteristics of complex systems.  Another advantage of the EOM framework is its general character and the possibility to accommodate various truncation schemes. For example, the simplest truncation on one-body level leads to the Hartree-Fock, random phase approximation (RPA), Second RPA (SRPA), the Gorkov theory of the superfluidity and the Bardeen-Cooper-Schrieffer (BCS) model. Explicit inclusion of higher-rank propagators leads to more complicated sets of coupled equations for propagators of different ranks.
A considerable accuracy can be achieved by cluster expansions of the dynamical kernels of the fermionic EOM's in terms of the two-time many-fermion correlation functions corresponding to the relevant degrees of freedom, as it is discussed, in particular, in Refs. \cite{SchuckTohyama2016,Olevano2018}. 
An attractive feature of the formally exact EOM's for these correlation functions, which are known in condensed matter and quantum chemistry \cite{Tiago2008,Martinez2010,Sangalli2011,Olevano2018}, is that they have both the static and dynamical kernels derived consistently from the same underlying bare interaction. In nuclear physics, however, the implementations of the analogous methods are still mainly based on phenomenological interactions. 

The formal foundation of such methods typically postulate phenomenological Hamiltonians, which imply the existence of the fermionic quasiparticles and phonons of bosonic nature.  The phonon-exchange interaction between the quasiparticles is added to the pure effective residual interaction between fermions, for instance, in the nuclear field theory (NFT) \cite{BohrMottelson1969,BohrMottelson1975,Broglia1976,BortignonBrogliaBesEtAl1977,BertschBortignonBroglia1983}. Another class of models is based on the phonon degrees of freedom \cite{Soloviev1992,Ponomarev2001,SavranBabilonBergEtAl2006,Andreozzi2008}. The use of effective phenomenological interactions allows for simpler calculation schemes, however, more accurate and sophisticated versions of the nuclear field theory (NFT) were successfully implemented \cite{IdiniBarrancoVigezzi2012,Litvinova2016,RobinLitvinova2016,RobinLitvinova2018,Niu2018,Robin2019,Tselyaev2018,Lyutorovich:2018cph,LitvinovaSchuck2019,Shen2020}. Analogous models operating with mostly the phonon degrees of freedom \cite{Soloviev1992,Ponomarev2001,SavranBabilonBergEtAl2006,Andreozzi2008}, were extended to very complex correlations, and there have been a few recent attempts of using the bare nucleon-nucleon interaction \cite{DeGregorio2017,DeGregorio2017a,Knapp:2014xja,Knapp:2015wpt}.  

In this work we present a finite-temperature extension of the many-body model for the nuclear shell structure \cite{LitvinovaRing2006,Litvinova2012}, which combines the relativistic quantum hadrodynamics (QHD) \cite{SerotWalecka1986a,SerotWalecka1979,Ring1996,VretenarAfanasjevLalazissisEtAl2005} and the quantum field theory techniques based on the EOM \cite{RingSchuck1980,DukelskyRoepkeSchuck1998,LitvinovaSchuck2019,Schuck2019}. We focus on the single-particle EOM, which is represented by the Dyson equation, and elaborate on its dynamical kernel. As it follows from Refs. \cite{DukelskyRoepkeSchuck1998,LitvinovaSchuck2019}, the leading contribution to the dynamical kernel, also called self-energy,  
in finite nuclei can be associated with the particle-vibration coupling (PVC). Formally similar to the phenomenological PVC proposed quite early by A. Bohr and B. Mottelson \cite{BohrMottelson1969,BohrMottelson1975} and that of the NFT, it is now understood in terms of the EOM derived from ab-initio nucleon-nucleon potentials. Although in this work we still keep the phenomenological effective interaction adjusted in the framework of the covariant density functional theory (CDFT) \cite{VretenarAfanasjevLalazissisEtAl2005} for the static part of the EOM kernel and PVC, it is supposed to pave the way to a fully ab initio description in near future.

The approach is designed to clarify the nuclear phenomena, which may occur in astrophysical environments, such as neutron stars and supernovae. In such environments finite temperature becomes an essential factor, which modifies the rates of various nuclear reactions, from the radiative neutron capture to the weak processes \cite{LitvinovaWibowo2018,LitvinovaRobinWibowo2020}. It was pointed out  in earlier works, such as Refs. \cite{Donati1994,Fantina2011}, that the nuclear single-particle states underly the mechanisms of those processes and impact the statistical properties, such as the level density, entropy and specific heat. At the same time, the single-particle states are modified considerably by the PVC mechanism, that results into an enhancement of the effective mass and of the level density around the Fermi surface. The results presented in this work allow for a detailed discussion of the latter phenomenon and its evolution with temperature.  

\section{Dyson equation for the fermionic propagator at finite temperature}
\label{Setup}

We define the atomic nucleus as a many-body quantum system, which consists of protons and neutrons, commonly called nucleons, in the regimes associated with the energies below the pion mass. Nucleons are coupled through the strong interaction represented by the meson exchange at such low energies. The Coulomb interaction acts between the positively charged protons. In this work we will use the concept of effective mesons, whose masses and coupling vertices are adjusted to reproduce the nuclear masses and radii on the Hartree level.
While the latter constitutes the famous Walecka model for nuclear QHD, it will be used, as in the preceding series of works, only as a starting point and a convenient basis for the description of nucleonic in-medium correlations far beyond the mean field. 

One of the most convenient ways to quantitatively approach these correlations is to directly calculate certain correlation functions, namely the propagators (also called Green functions). We will employ the advantage of this formalism as it gives a direct access to the excitation spectra and ground state properties of the nuclear system. Here we are interested in nuclear systems, which are in thermal equilibrium with the surroundings and can thus be assigned the temperature. The temperature, or Matsubara, Green function of a fermion is defined as \cite{Matsubara1955, Abrikosov1975, Zagoskin2014}
\be
{\cal G}(1,1') \equiv {\cal G}_{k_1k_{1'}}(\tau_1-\tau_{1'}) = -\langle T_{\tau} \psi(1){\bar\psi}(1') \rangle,
\label{spgf}
\ee
with the help of the chronological ordering operator $T_{\tau}$, which acts on the fermionic field operators in the Wick-rotated picture:
\bea
\psi(1) &\equiv& \psi_{k_1}(\tau_1) = e^{{\cal H}\tau_1}\psi_{k_1}e^{-{\cal H}\tau_1},\nonumber\\
{\bar\psi}(1) &\equiv& {\psi^{\dagger}}_{k_1}(\tau_1) = e^{{\cal H}\tau_1}{\psi^{\dagger}}_{k_1}e^{-{\cal H}\tau_1},
\label{Wick-Heisenberg}
\eea
where ${\cal H} = H - \mu N$ with $H$ being the many-body Hamiltonian, $\mu$ the chemical potential, and $N$ the particle number operator.
The subscript $k_1$ stands for the full set of the single-particle quantum numbers in a given representation and the imaginary time variables $\tau$ are related to the real times $t$ as  $\tau = it$. The fermionic fields satisfy the usual anticommutation relations, and the angular brackets in Eq. (\ref{spgf}) stand for the thermal average \cite{Abrikosov1975, Zagoskin2014}.

If the many-body Hamiltonian $H$ is confined by the one-body part, i.e. contains only the free-motion and the mean-field contributions, the single-fermion Matsubara Green function can be easily calculated and reads
\bea
\widetilde{\cal G}(2,1) &=& \sum\limits_{\sigma=\pm1}\widetilde{\cal G}^{\sigma}(2,1),\nonumber\\
\widetilde{\cal G}^{\sigma}(2,1) &=& -\sigma\delta_{k_{2}k_{1}}n(-\sigma(\varepsilon_{k_1}-\mu),T)e^{-(\varepsilon_{k_{1}}-\mu)\tau_{21}}\theta(\sigma\tau_{21})\nonumber\\
\label{mfgf}
\eea
with $\tau_{21} = \tau_{2} - \tau_{1}$ in the basis $\{k_i\}$ of the single-fermion states, which diagonalizes the one-body part of the Hamiltonian. In Eq. (\ref{mfgf})
$\varepsilon_{k_1}$ are the eigenvalues of the single-particle part of the Hamiltonian and $n(\varepsilon,T)$ stands for the Fermi-Dirac distribution 
\be
n(\varepsilon,T) = \frac{1}{\text{exp}(\varepsilon/T) + 1} 
\label{FD}
\ee
at the temperature $T$.
The spectral representation of the thermal mean-field Green function is calculated with the operation
\be
\widetilde{{\cal G}}_{k_{2}k_{1}}(\varepsilon_{\ell}) = \int\limits_{0}^{1/T}d \tau e^{i\varepsilon_{\ell}\tau} \widetilde{\cal G}_{k_2k_1}(\tau),
\ee 
which leads to:
\be
\widetilde{{\cal G}}_{k_{2}k_{1}}(\varepsilon_{\ell})=\delta_{k_{2}k_{1}}\widetilde{{\cal G}}_{k_{1}}(\varepsilon_{\ell}),\;\;\;\;\;\;\;\;\;\;\widetilde{{\cal G}}_{k_{1}}(\varepsilon_{\ell})=\frac{1}{i\varepsilon_{\ell}-\varepsilon_{k_{1}}+\mu}
\label{mfgfs}
\ee
defined at the discrete Matsubara frequencies 
$\varepsilon_{\ell}$ 
\begin{eqnarray}
\varepsilon_{\ell}=(2\ell+1)\pi T,
\end{eqnarray}
where the $\ell$'s are integer. The '$\widetilde{\ \ }$' sign in Eqs. (\ref{mfgf}-\ref{mfgfs}) indicates the mean-field character of the respective Green function.

In this work we are interested in non-trivial correlations beyond mean field which occur due to the residual interaction, i.e. in the presence of two-body and higher-rank terms in the many-body Hamiltonian. In this case, the single-fermion propagator $\cal G$ obeys the Dyson equation 
\begin{equation}
\label{Dyson Equation for Exact One-Body Temperature Green's Function}{\cal G}_{k_{1}k_{2}}(\varepsilon_{\ell})={\cal G}^{0}_{k_{1}k_{2}}(\varepsilon_{\ell})+\sum_{k_{3}k_{4}}{\cal G}^{0}_{k_{1}k_{3}}(\varepsilon_{\ell})\Sigma_{k_{3}k_{4}}(\varepsilon_{\ell}){\cal G}_{k_{4}k_{2}}(\varepsilon_{\ell}),
\end{equation}
where ${\cal G}^{0}$ is the free propagator and $\Sigma$ is the self-energy, or the mass operator. As it can be shown within the equation of motion (EOM) framework, the exact self-energy is decomposed into the energy-independent (static) $\widetilde{\Sigma}$ and the energy-dependent (dynamical)  $\Sigma^{e}$ parts:
\begin{equation}
\Sigma_{k_{3}k_{4}}(\varepsilon_{\ell})=\widetilde{\Sigma}_{k_{3}k_{4}}+\Sigma^{e}_{k_{3}k_{4}}(\varepsilon_{\ell}),
\label{SE}
\end{equation}
that is also valid at finite temperature. In 'ab-initio' calculations based on the Hamiltonians with bare interactions the static part of the self-energy is given by the contraction of the matrix element of the bare interaction with the exact one-fermion density, while its dynamical part is represented by the three-fermion correlated propagator contracted with two matrix elements  of the bare interaction \cite{AdachiSchuck1989, DukelskyRoepkeSchuck1998,LitvinovaSchuck2019}. 


Using Eq. (\ref{SE}), it is convenient to eliminate the unperturbed propagator ${\cal G}^{0}$ from Eq. \eqref{Dyson Equation for Exact One-Body Temperature Green's Function}, and work with the thermal mean-field propagator $\widetilde{{\cal G}}$ which satisfies the equation:
\begin{equation}
\label{DysonMF}\widetilde{{\cal G}}_{k_{1}k_{2}}(\varepsilon_{\ell})={\cal G}^{0}_{k_{1}k_{2}}(\varepsilon_{\ell})+\sum_{k_{3}k_{4}}{\cal G}^{0}_{k_{1}k_{3}}(\varepsilon_{\ell})\widetilde{\Sigma}_{k_{3}k_{4}}\widetilde{{\cal G}}_{k_{4}k_{2}}(\varepsilon_{\ell}).
\end{equation}
Then, the Dyson equation for the full propagator takes the form:
\begin{equation}
\label{Dyson}{\cal G}_{k_{1}k_{2}}(\varepsilon_{\ell})=\widetilde{{\cal G}}_{k_{1}k_{2}}(\varepsilon_{\ell})+\sum_{k_{3}k_{4}}\widetilde{{\cal G}}_{k_{1}k_{3}}(\varepsilon_{\ell})\Sigma^{e}_{k_{3}k_{4}}(\varepsilon_{\ell}){\cal G}_{k_{4}k_{2}}(\varepsilon_{\ell}).
\end{equation}
The energy-dependent part of the mass operator $\Sigma^{e}$ describes the coupling between single fermions and in-medium emergent degrees of freedom. In this work, we employ the particle-vibration coupling model, which approximates the exact energy-dependent part of the self-energy $\Sigma^{e}$ by a cluster expansion truncated at the two-body level \cite{LitvinovaSchuck2019}. When retaining only the coupling to normal phonons, the analytical form of this self-energy, in the leading approximation, reads
\begin{equation}
\Sigma^{e}_{k_{1}k_{2}}(\varepsilon_{\ell})=-T\sum_{k_{3},m}\sum_{\ell'}\sum_{\sigma=\pm 1}\widetilde{{\cal G}}_{k_{3}}(\varepsilon_{\ell'})\frac{\sigma g^{m(\sigma)}_{k_{1}k_{3}}g_{k_{2}k_{3}}^{m(\sigma)\ast}}{i\varepsilon_{\ell}-i\varepsilon_{\ell'}-\sigma\omega_{m}}
\label{SE1}
\end{equation}
where $g^{m}$ are the phonon vertices and $\omega_m$ their frequencies, which can be found from the EOM for the two-fermion correlation functions. The phonon vertices are determined via
\begin{eqnarray}
g^{m}_{k_{1}k_{2}}=\sum_{k_{3}k_{4}}\widetilde{\mathscr{U}}_{k_{1}k_{4},k_{2}k_{3}}\rho^{m}_{k_{3}k_{4}}, \label{vertex}\\
g^{m(\sigma)}_{k_{1}k_{2}} = \delta_{\sigma,+1}g^{m}_{k_{1}k_{2}} + \delta_{\sigma,-1}g^{m\ast}_{k_{2}k_{1}}
\end{eqnarray}
where $\rho^{m}_{k_{3}k_{4}}$ are the transition densities of the phonons and $\widetilde{\mathscr{U}}_{k_{1}k_{4},k_{2}k_{3}}$ are the matrix elements of the nucleon-nucleon interaction. In principle, the relationship (\ref{vertex}) is model-independent and, ideally, the transition densities are the exact ones, while the interaction $\widetilde{\mathscr{U}}$ is the bare interaction. However, numerous models employing effective interactions and the random phase approximation based on these interactions for the computation of the phonon vertices and frequencies typically provide quite realistic approaches to the dynamical self-energy.
In this work, we use the effective interaction of the covariant energy density functional (CEDF) \cite{Ring1996,VretenarAfanasjevLalazissisEtAl2005} with the NL3 parametrization \cite{Lalazissis1997} and the concept of the 'no-sea' relativistic random phase approximation (RRPA) \cite{RingMaVanGiaiEtAl2001} adopted to finite temperature in our previous developments \cite{LitvinovaWibowo2018,WibowoLitvinova2019,LitvinovaWibowo2019}.

The summation over $\ell'$ in Eq. (\ref{SE1}) can be transformed into a contour integral by the standard technique \cite{Zagoskin2014}. After the analytical continuation to  complex energies, we then obtain the final expression for the mass operator $\Sigma^{e}$ of the form:
\begin{eqnarray}
{\Sigma}^{e}_{k_{1}k_{2}}(\varepsilon)&=&\sum_{k_{3},m}\Bigg\{g^{m}_{k_{1}k_{3}}g^{m\ast}_{k_{2}k_{3}}\frac{N(\omega_{m},T)+1-n(\varepsilon_{k_{3}}-\mu,T)}{\varepsilon-\varepsilon_{k_{3}} + \mu -\omega_{m}+i\delta}\nonumber\\ 
&+& g^{m\ast}_{k_{3}k_{1}}g^{m}_{k_{3}k_{2}}\frac{n(\varepsilon_{k_{3}}-\mu,T)+N(\omega_{m},T)}{\varepsilon-\varepsilon_{k_{3}} + \mu+\omega_{m}-i\delta}\Bigg\},\nonumber\\
\label{SET}
\end{eqnarray}
where 
\begin{equation}
N(\omega_{m},T)=\frac{1}{\text{exp}({\omega_{m}/T})-1}
\end{equation}
are the occupation numbers of phonons with the frequencies $\omega_{m}$. 
It is easy to see that in the limit $T\rightarrow 0$ Eq. \eqref{SET} recovers the zero-temperature result for the energy-dependent mass operator
in the PVC model:
\bea
\Sigma^{e}_{k_{1},k_{2}}(\varepsilon)=\sum_{\substack{k_{3},m \\ {\varepsilon}_{k_{3}}>\varepsilon_{F}}}\frac{g^{m}_{k_{1}k_{3}}g^{m\ast}_{k_{2}k_{3}}}{\varepsilon-{\varepsilon}_{k_{3}}+ \varepsilon_F-\omega_{m}+i\delta}\nonumber\\
+\sum_{\substack{k_{3},m \\ {\varepsilon}_{k_{3}}\leq\varepsilon_{F}}}\frac{g^{m\ast}_{k_{3}k_{1}}g^{m}_{k_{3}k_{2}}}{\varepsilon-{\varepsilon}_{k_{3}}+ \varepsilon_F+\omega_{m}-i\delta},\;\;\;\;\;\delta\rightarrow+0,
\label{Mass Operator at T = 0 (2)}
\eea
where $\varepsilon_F$ is the Fermi energy, i.e. the energy of the last occupied single-particle state. 
Indeed, in the limit $T\rightarrow 0$ the phonon occupation number $N(\omega_{m},T)\rightarrow 0$, and the fermion occupation number $n(\varepsilon_{k_{3}}-\mu,T)$ takes the value 1 (0) for $\varepsilon_{k_{3}}\leq\varepsilon_{F}$ $(\varepsilon_{k_{3}}>\varepsilon_{F})$.


\section{Numerical solution of the finite-temperature Dyson equation: results and discussion}

We have selected the atomic nuclei $^{56,68}$Ni and $^{56}$Fe to illustrate the performance of the developed approach. This choice was determined, in particular, by the astrophysical relevance of these nuclear systems. Indeed, nickel and iron isotopes with the mass number $A = 56$ play a very important role in current understanding of stellar evolution. For instance, they are associated with the final stage of formation of massive evolved stars before core collapse. Characteristics of all the three nuclei are important ingredients for understanding presupernova and neutron stars. Also, comparison between $^{56}$N and $^{68}$Ni can be informative for evaluating the range of the relevant characteristics within a single isotopic chain. 

The calculation scheme consists of the following steps. First, we solve the closed set of the relativistic mean field (RMF) equations with the NL3 parametrization \cite{Lalazissis1997} of the non-linear sigma-model with the thermal fermionic occupation numbers (\ref{FD}). This leads to a set of temperature-dependent single-particle Dirac spinors and the corresponding single-nucleon energies, which form the basis for subsequent calculations. 
Second, the finite-temperature relativistic random phase approximation (FT-RRPA) equations are solved to obtain the phonon vertices $g^{m}$ and their frequencies $\omega_{m}$. 
The set of the obtained FT-RRPA phonons, 
together with the RMF single-nucleon basis, forms the $pp\otimes$phonon and $ph\otimes$phonon configurations for the particle-phonon coupling self-energy $\Sigma^{e}(\varepsilon)$.
Third,  Eq. (\ref{Dyson}) is solved in the truncated configuration space, as described in \cite{LitvinovaRing2006} for the $T=0$ case, in the diagonal approximation, i.e. $\Sigma^{e}_{k_1k_2}(\varepsilon) = \delta_{k_1k_2}\Sigma^{e}_{k_1}(\varepsilon)$. 

The particle-hole basis for the FT-RRPA calculations of the phonons was limited by the particle-hole ($ph$) configurations with the energies $\varepsilon_{ph}\leq100$ MeV and the antiparticle-hole ($\alpha h$) ones with $\varepsilon_{\alpha h}\geq-1800$ MeV with respect to the positive-energy continuum. Test calculations within the complete RMF basis showed that the excitation spectra converge reasonably well with this truncation. The set of phonons included vibrations with the quantum numbers of spin and parity $J^{\pi}=2^{+},\;3^{-},\;4^{+},\;5^{-},\;6^{+}$ below the energy cutoff, which amounts to 20 MeV for the considered nuclei. This cutoff is justified by our previous calculations. A truncation of the phonon space was applied according to the values of the reduced transition probabilities of the corresponding electromagnetic transitions: we included the modes with the reduced transition probabilities $B(EL)$ equal or more than 5\% of the maximal one (for each $J^{\pi}$). The same truncation criteria are applied to the phonon energy, $J^{\pi}$ and the reduced transition probability for all temperature regimes in order to make a fair comparison of the calculated single-particle strength distributions at different temperatures. As in our previous calculations \cite{LitvinovaWibowo2018,WibowoLitvinova2019,LitvinovaWibowo2019}, at high temperatures we see the appearance of many additional phonon modes as a consequence of the significant thermal unblocking, that may cause slower saturation of the results with respect to the $B(EL)$ cutoff. This happened typically at the temperatures of 5-6 MeV, which we do not consider here because of their little relevance to the astrophysical applications. Another truncation was made on the 
single-particle intermediate states $k_3$ in the summation of Eq. (\ref{SET}): only the states with the energy differences  $|\varepsilon_{k_3}- \varepsilon_{k_1}| \leq$ 50 MeV were included in the summation. 
\begin{figure}
\begin{center}
\includegraphics[scale=0.35]{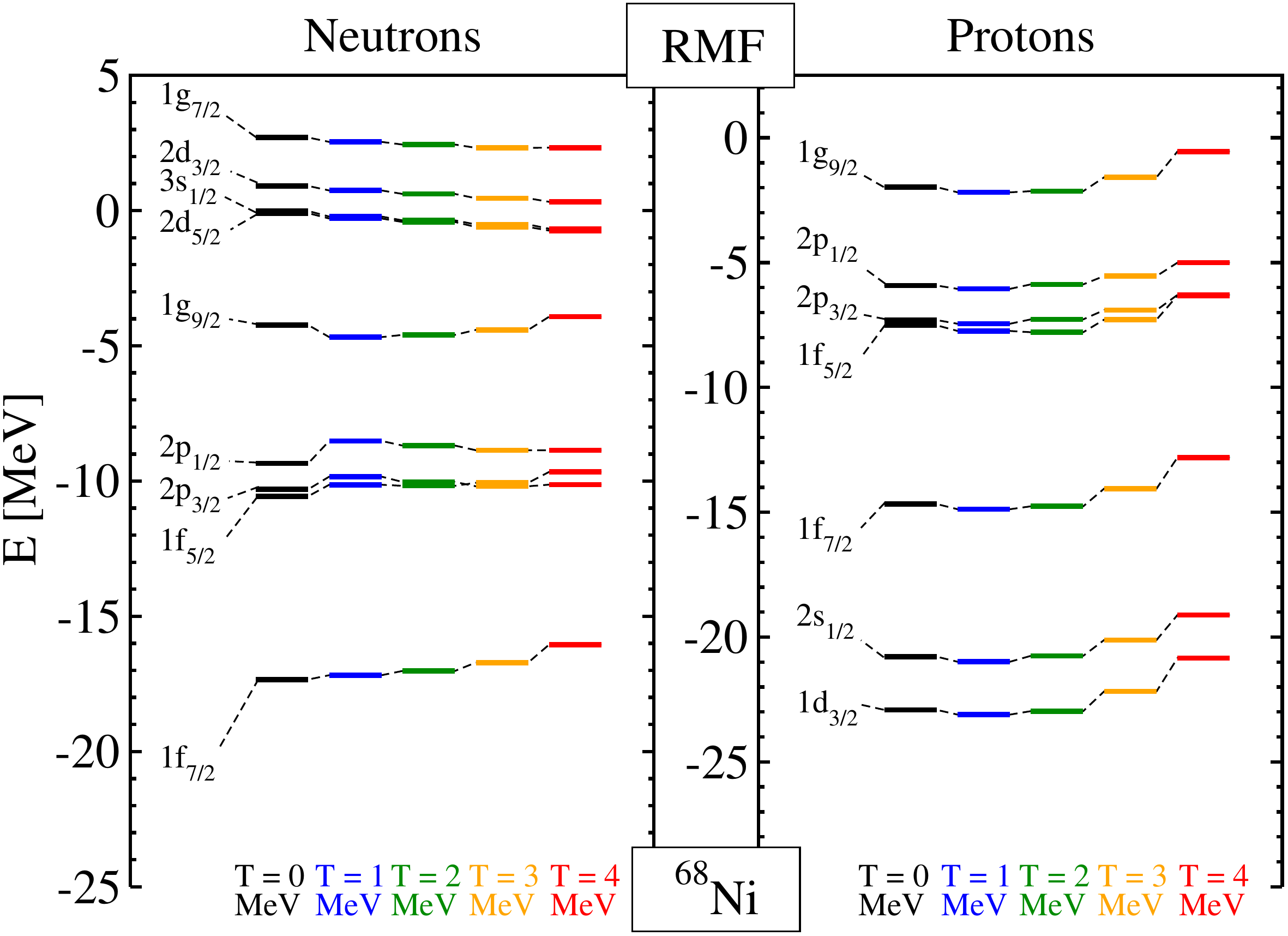}
\end{center}
\caption{Single-particle states in $^{68}$Ni at zero and finite temperatures calculated in the RMF approximation.}
\label{RMF}%
\end{figure}
\begin{figure}
\begin{center}
\includegraphics[scale=0.35]{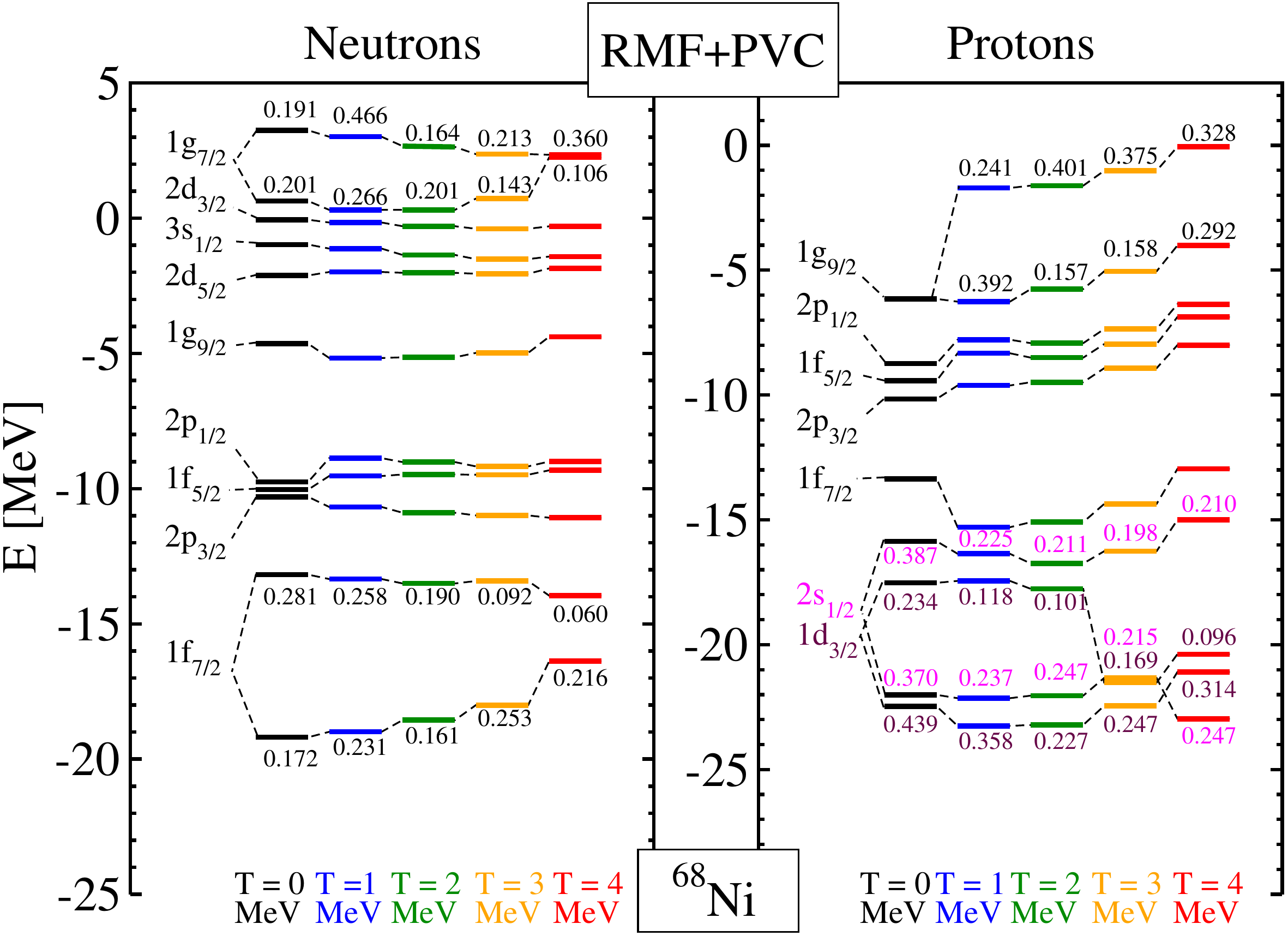}
\end{center}
\caption{The dominant fragments of the single-particle states in $^{68}$Ni at zero and finite temperatures calculated in the RMF+PVC approximation.}
\label{RMF-PVC}%
\end{figure}

We investigated the neutron and proton states in the approximately 20 MeV energy window around the respective Fermi energies of $^{56,68}$Ni and $^{56}$Fe nuclei. While $^{56}$Ni is a doubly-magic, or closed-shell, nucleus, the proton subsystem of $^{68}$Ni is of the closed-shell nature and the neutron subsystem is open-shell. The $^{56}$Fe is, in turn, an open-shell nucleus for both neutrons and protons. Thus, the superfluid character of the respective subsystems, because of the presence of the pairing correlations, can importantly affect the single-particle spectra. Within our formalism, this phenomenon was discussed in Refs. \cite{Litvinova2012,AfanasjevLitvinova2015}. On the mean-field level, the Bardeen-Cooper-Schrieffer (BCS) or the Bogoliubov's approximations typically give a reasonable description of pairing correlations, which manifest themselves through considerable redistributions of the single-particle states in the vicinity of the Fermi energy and their fractional occupancies in the superfluid subsystems, along with some minor rearrangements of the single-particle states in their non-superfluid counterparts. In the calculations beyond the mean field, such as RMF+PVC, pairing correlations may have stronger impacts. The main underlying reason for these impacts is that taking pairing correlations into account leads to the appearance of the phonon modes, first of all, the quadrupole modes, at significantly lower energies. Such phonons play the major role in the dynamical self-energy. As it has been shown in Ref. \cite{Litvinova2012}, the modification of the phonon spectrum due to the pairing correlations further affects the single-particle structure, as compared to the mean-field approach. The PVC fragmentation effects become stronger in general and, in particular, the shell structure of the non-superfluid counterparts are modified considerably toward much higher single-particle level densities.  

As the superfluidity in the Bogoliubov's or BCS sense vanishes at the critical temperature $T_c$, which has a well-established relation to the pairing gap $\Delta_p$ at zero temperature:  $T_c \approx 0.6\Delta_p(T=0)$, the role of pairing correlations diminishes quickly with the temperature growth. For the nuclei considered in this work, whose pairing gaps were adjusted to the odd-even mass differences using the three-point formula and the data on nuclear binding energies from Ref. \cite{AudiWapstraThibault2003}, the values of pairing gaps deduced by this procedure are $\Delta^{(n)}_p$ = 1.6 MeV for neutrons in $^{68}$Ni, and $\Delta^{(n)}_p$ = 1.8 MeV and $\Delta^{(p)}_p$ = 2.1 MeV for neutrons and protons in $^{56}$Fe, respectively. For these cases the critical 
temperatures have the values in the 0 $\leq T \leq$ 1.3 MeV interval. Therefore, in the results presented below on 1 MeV temperature grid pairing correlations in superfluid systems are taken into account for $T = 0$ and neglected for $T \geq$ 1 MeV.  Accordingly, a transition from superfluid to non-superfluid phases between $T = 0$ and $T = 1$ MeV takes place in $^{68}$Ni. In $^{56}$Fe, where the neutron pairing gap $\Delta^{(n)}_p$ vanishes at approximately 1.1 MeV and the proton pairing gap $\Delta^{(p)}_p$ at approximately 1.3 MeV,  we assume that the role of pairing correlations is negligible already at $T = 1$ MeV, too. The $T = 0$ calculations with pairing correlations are performed within the superfluid RMF+PVC approach developed originally in Ref. \cite{Litvinova2012}. 

The obtained single-particle shell structure for $^{68}$Ni is shown in Figs. \ref{RMF} and \ref{RMF-PVC}. Fig. \ref{RMF} displays the states computed within the thermal RMF approach, while Fig.  \ref{RMF-PVC} presents the dominant states, which are the outcome of calculations within the finite-temperature particle-vibration coupling (RMF+PVC).
It is quite common in the literature to assume that the nuclear mean field remains almost unchanged with temperature in a relatively broad temperature range $0\leq T \leq 3$ MeV, and this assumption is widely used in the finite-temperature calculations of the nuclear shell structure and response \cite{Donati1994,Dzhioev2015,Dzhioev2016}. However, as one can see from 
Fig. \ref{RMF}, the mean-field states can develop quite sizably within these temperatures. In addition, the chemical potential has some temperature dependence. These effects had, for instance, a notable influence on the spin-isospin response, which was investigated within our framework in Ref. \cite{LitvinovaRobinWibowo2020}. While in the latter study this effect mixed with the thermal unblocking, in the present work the temperature evolution of the nuclear mean-field characteristics can be tracked explicitly. Besides the superfluid phase transition at about $T = 1$ MeV, the overall trend in the neutron subsystem is the relatively minor densifying of the spectrum with the temperature increase, while the proton mean-field states move up nearly uniformly by 1-2 MeV in the 0$\leq T \leq$ 4 MeV temperature interval.  

The dominant states shown in Fig. \ref{RMF-PVC} are the states obtained as a full solution of Eq. (\ref{Dyson}) with the maximal spectroscopic factors for each spherical single-particle quantum number set. As it is known from numerous studies of the PVC effects on the nuclear single-particle states, including the relativistic ones \cite{LitvinovaRing2006,Litvinova2012, LitvinovaAfanasjev2011,AfanasjevLitvinova2015}, the typical outcome of such studies is the fragmentation of the single-particle states as compared to the mean-field ones. However, the fragmentation due to the PVC mechanism is of a selective character. In general, if the system has closed shells or subshells, like in the case of $^{68}$Ni, for the each state close to the Fermi energy there is one fragment with the spectroscopic factor of 0.8-0.9, while the rest of the fragments are characterized by very little, mostly less than one percent spectroscopic factors. The fragment with the largest spectroscopic factor is commonly called dominant, and the states with this fragmentation pattern are called good single-particle states. Typically, for such states the energy of the dominant fragment is rather close to the energy of the original mean-field state. In other words, in non-superfluid systems the single-particle states in the vicinity of the Fermi energy are not very much affected by the PVC.  In contrast, the states far away from the Fermi energy are strongly fragmented. For many states it is still possible to identify the dominant fragment, although with a considerably quenched spectroscopic factor, and this fragment can be quite far from the original mean-field state. The algebraic reasons behind this qualitatively outlined picture are discussed, for instance, in Ref. \cite{LitvinovaRing2006}.  

For the second type of states, i.e. the states remote from the Fermi energy, often two or more fragments exhibit comparable spectroscopic factors. These are the cases, for instance, for the 1f$_{7/2}$ state in the neutron subsystem and 2s$_{1/2}$ state in the proton subsystem, which show 0.28/0.17 and 0.39/0.37 shares of the spectroscopic factors between two dominant fragments, respectively, at $T = 0$. Our calculations reveal that the general pattern described above persists with the temperature increase within the interval 0$\leq T \leq$ 4 MeV: although the entire spectrum exhibits quite a remarkable evolution, the states in the vicinity of the Fermi energy remain good single-particle states, while those away from the Fermi energy remain strongly fragmented. In particular, as one can observe from Fig. \ref{RMF-PVC}, the states  $\nu$1f$_{7/2}$, $\nu$1g$_{7/2}$, $\pi$2s$_{1/2}$ and  $\pi$1g$_{9/2}$ show splitting into two major fragments with comparable strengths. The gaps between these fragments as well as the ratios of their spectroscopic factors change with the temperature increase.

\begin{figure}
\begin{center}
\includegraphics[scale=0.35]{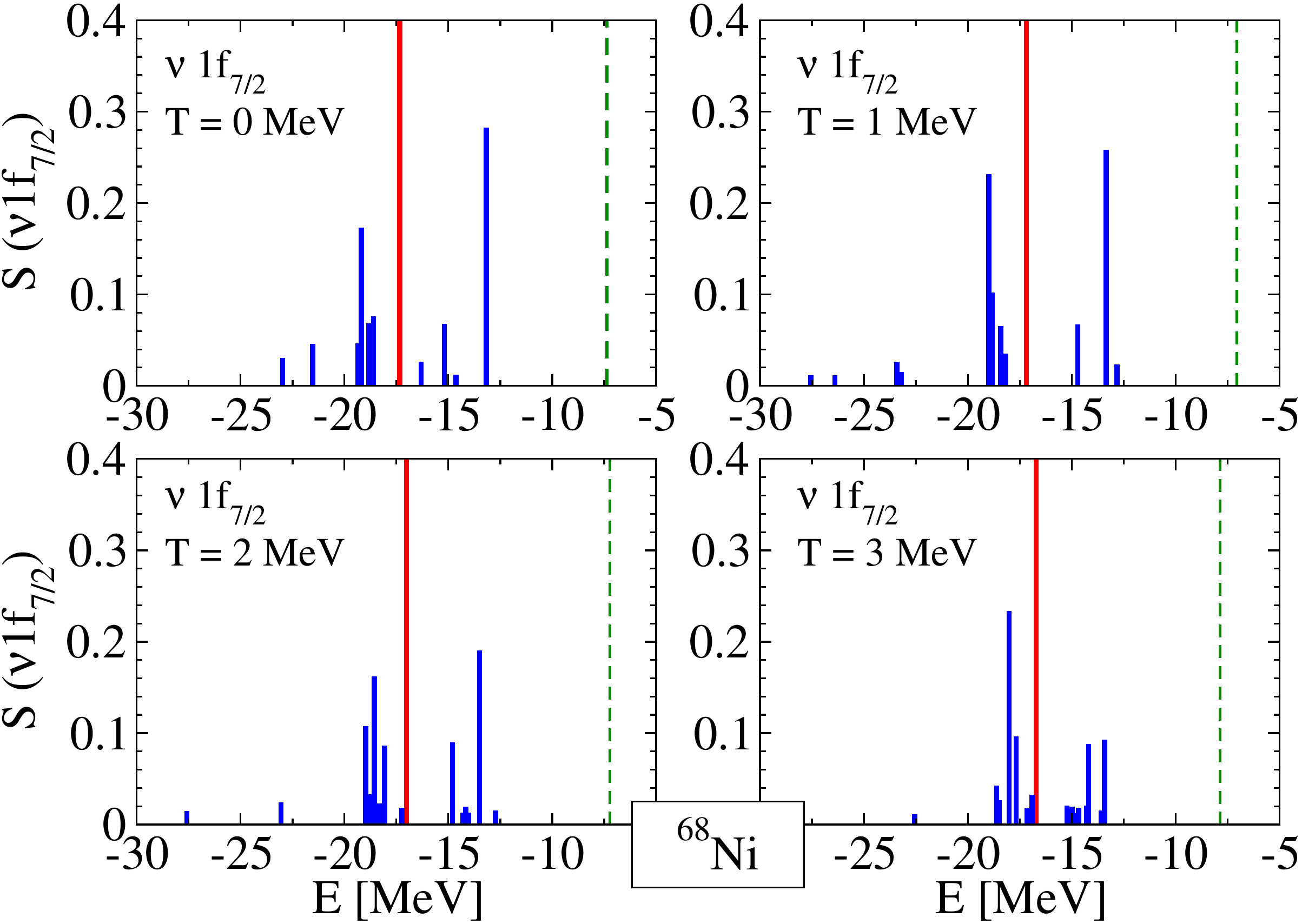}
\end{center}
\caption{Temperature evolution of the neutron 1f$_{7/2}$ state in $^{68}$Ni. Blue bars represent the spectral strength distributions (spectroscopic factors) of the  fragmented state calculated within the thermal 'RMF+PVC' approach. The red bar corresponds to the pure RMF state, and the dashed green line indicates the chemical potential.}
\label{68ni_nf72}%
\end{figure}
\begin{figure}
\begin{center}
\includegraphics[scale=0.35]{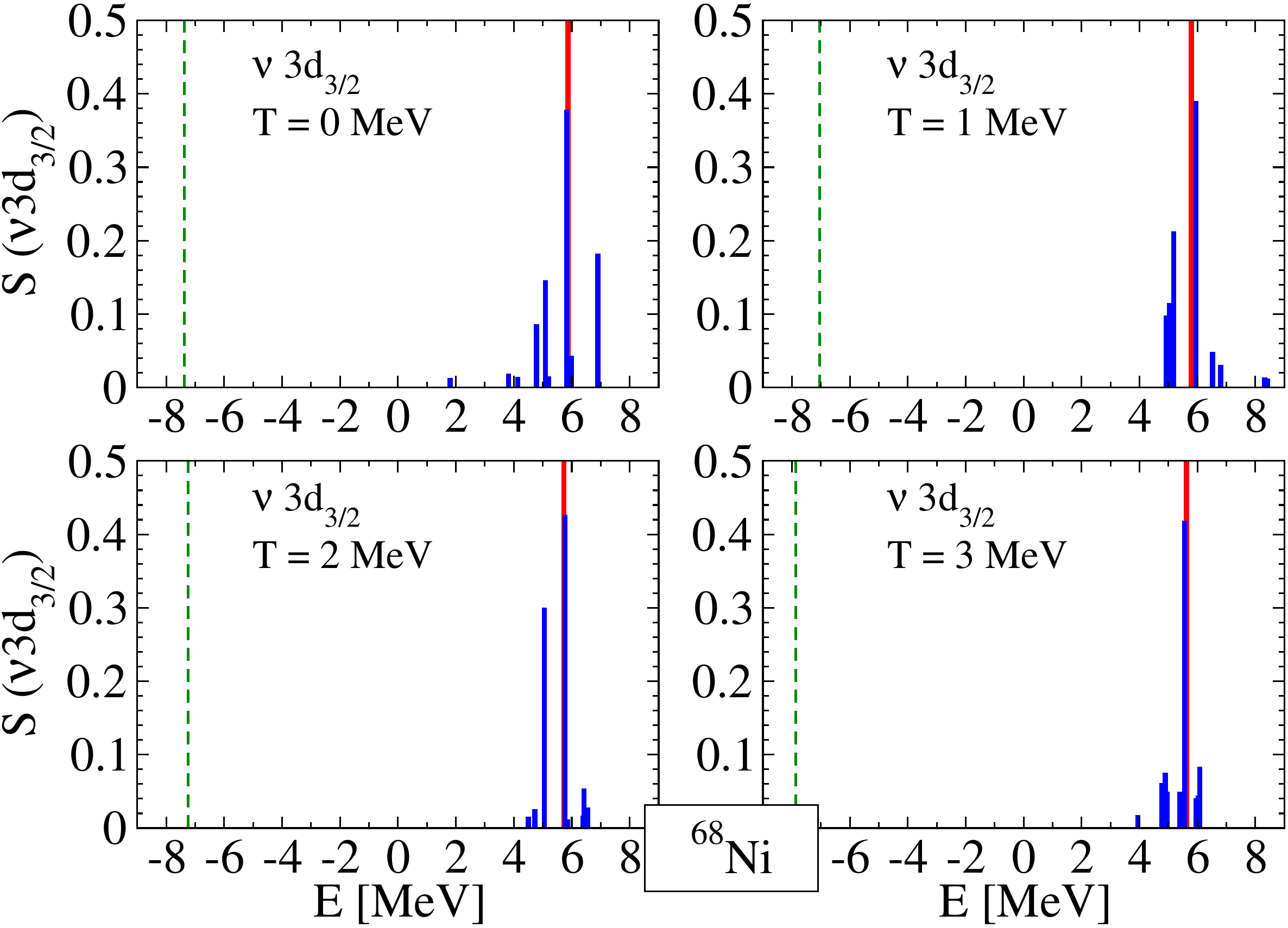}
\end{center}
\caption{Same as in Fig. \ref{68ni_nf72}, but for the neutron 3d$_{3/2}$ state in $^{68}$Ni.}
\label{68ni_nd32}%
\end{figure}

\begin{figure}
\begin{center}
\includegraphics[scale=0.35]{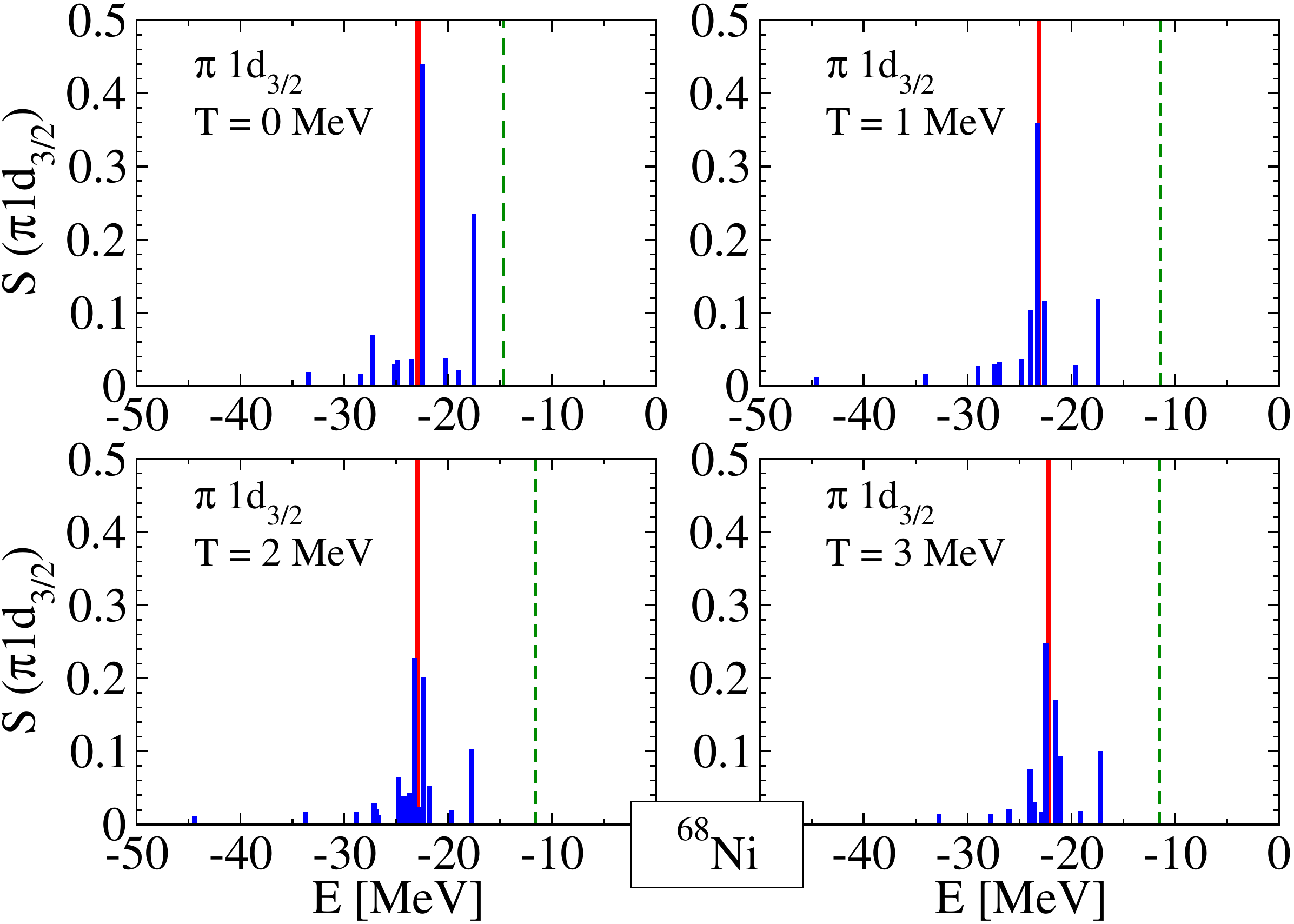}
\end{center}
\caption{Same as in Fig. \ref{68ni_nf72}, but for the proton 1d$_{3/2}$ state in $^{68}$Ni.}
\label{68ni_pd32}%
\end{figure}

\begin{figure}
\begin{center}
\includegraphics[scale=0.35]{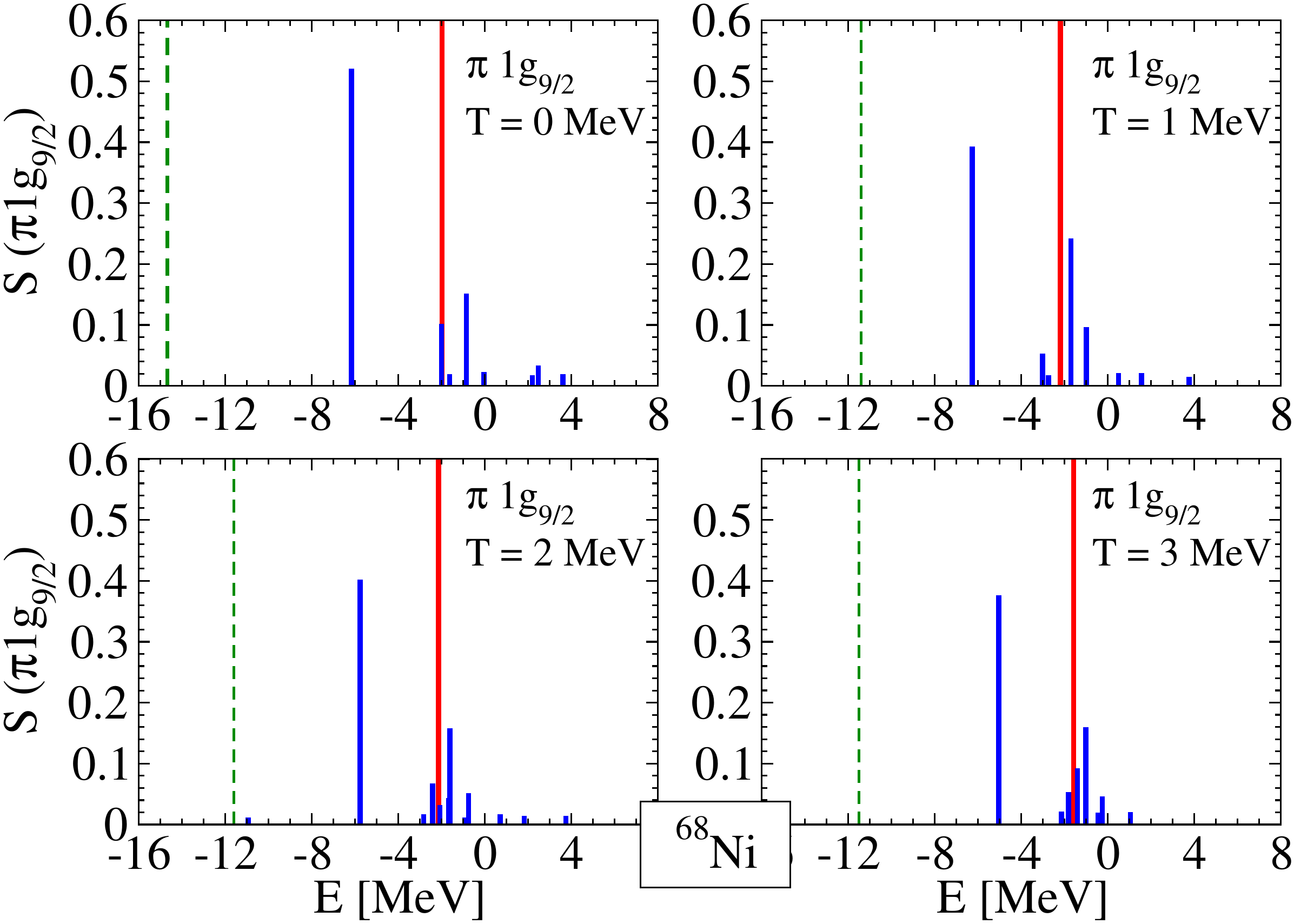}
\end{center}
\caption{Same as in Fig. \ref{68ni_nf72}, but for the proton 1g$_{9/2}$ state in $^{68}$Ni.}
\label{68ni_pg92}%
\end{figure}

In order to illustrate and better understand the temperature evolution of the fragmentation mechanism, we display four examples of strongly fragmented single-particle states in $^{68}$Ni at the temperatures of $T = 0, 1, 2$, and 3 MeV in Figs. \ref{68ni_nf72} - \ref{68ni_pg92}. The neutron state 1f$_{7/2}$ in $^{68}$Ni  is shown in Fig. \ref{68ni_nf72}.  At $T = 0$ it consists predominantly of the two fragments approximately 6 MeV apart  with the spectroscopic factors of 0.28 and 0.17 located on the opposite sides of the uncorrelated, or mean-field, hole state (below the Fermi energy). The phase transition, which occurs around T = 1 MeV, together with the beginning thermal unblocking, slightly changes the strength distribution preserving, however, the general two-peak structure. With further temperature increase, the two-peak structure persists, while each of the two peaks undergo fragmentation.  At temperatures T = 3 MeV and T = 4 MeV we find that the lower-energy major fragment dominates, although its spectroscopic factor continues to quench.  The evolution of the state  3d$_{3/2}$ in the neutron subsystem of $^{68}$Ni is illustrated in Fig. \ref{68ni_nd32}. This is the particle state (well above the Fermi energy), which appears at $T = 0$ as a structure with a single dominant peak and where at T = 1 another major fragment appears to compete with the share of 0.21/0.39 between the spectroscopic factors. In contrast to the case of the neutron 1f$_{7/2}$ state, these fragments are only about 1 MeV apart in energy and rather close to the mean-field 3d$_{3/2}$ state, from which they originate. With the temperature increase one of the two major fragments undergoes further fragmentation, and at T = 3 MeV the other fragment, which is closer to the original mean-field 3d$_{3/2}$ state, becomes absolutely dominant.

The examples from the proton subsystem are represented by the states 1d$_{3/2}$ and 1g$_{9/2}$ shown in Figs. \ref{68ni_pd32} and \ref{68ni_pg92}, respectively. The case of the proton 1d$_{3/2}$ state is similar to the one of the neutron 3d$_{3/2}$ state discussed above.  At $T = 0$ the dominant state appears surrounded by the multitude of weaker fragments and a relatively strong second fragments with 0.23/0.45 share of the spectroscopic factors.
With the temperature increase, the entire spectrum undergoes gradual fragmentation with strong quenching of the dominant fragment. In all temperature regimes the major fragment is located very close to the original mean-field state. The behavior of the proton 1g$_{9/2}$ state is again different. At $T = 0$ the dominant fragment splits out from a number of weaker fragments, which are grouped around the original mean-field state. After the phase transition, rearranging the distribution into a competing two-peak structure, the temperature growth causes a redistribution and eventually a stronger fragmentation of this group of fragments, which nevertheless remain around the mean-field state. The dominant fragment retains its dominance, although its spectroscopic factor gets gradually quenched as the temperature increases. 

Thereby, while the evolution of the good single-particle states is quite similar to the evolution of the mean-field states, the strongly fragmented states may exhibit various scenarios. The latter is determined by the spin and parity of the state, its closeness to the Fermi energy as well as by the magnitudes of the most relevant PVC matrix elements and the associated phonon frequencies. These factors define the interplay of coupling to various phonon modes \cite{Vaquero2020}. As we have shown explicitly for the case of the most important quadrupole modes in Ref. \cite{LitvinovaWibowo2019}, the non-trivial temperature evolution of the phonon spectrum gives the corresponding feedback on the PVC and the related fragmentation of the nuclear excitation modes. On the large scale of temperatures, the fragmentation is enhanced with the temperature  increase because of the general trend of the thermal unblocking. However, we also observed a counter trend at moderate temperatures, when the thermal unblocking is not yet developed sufficiently to generate new strong low-energy phonon modes, but weakens those modes which play the major role at $T = 0$.   
The feedback of this effect on the nuclear excited states was observed as a minor weakening of the fragmentation at small and moderate temperatures of about 1-2 MeV, while with further temperature increase the fragmentation became reinforced again. Now we can see that some of the single-particle states show a similar behavior: the examples of those are the neutron 3d$_{3/2}$ and the proton 1g$_{9/2}$ states. 

\begin{figure}
\begin{center}
\includegraphics[scale=0.35]{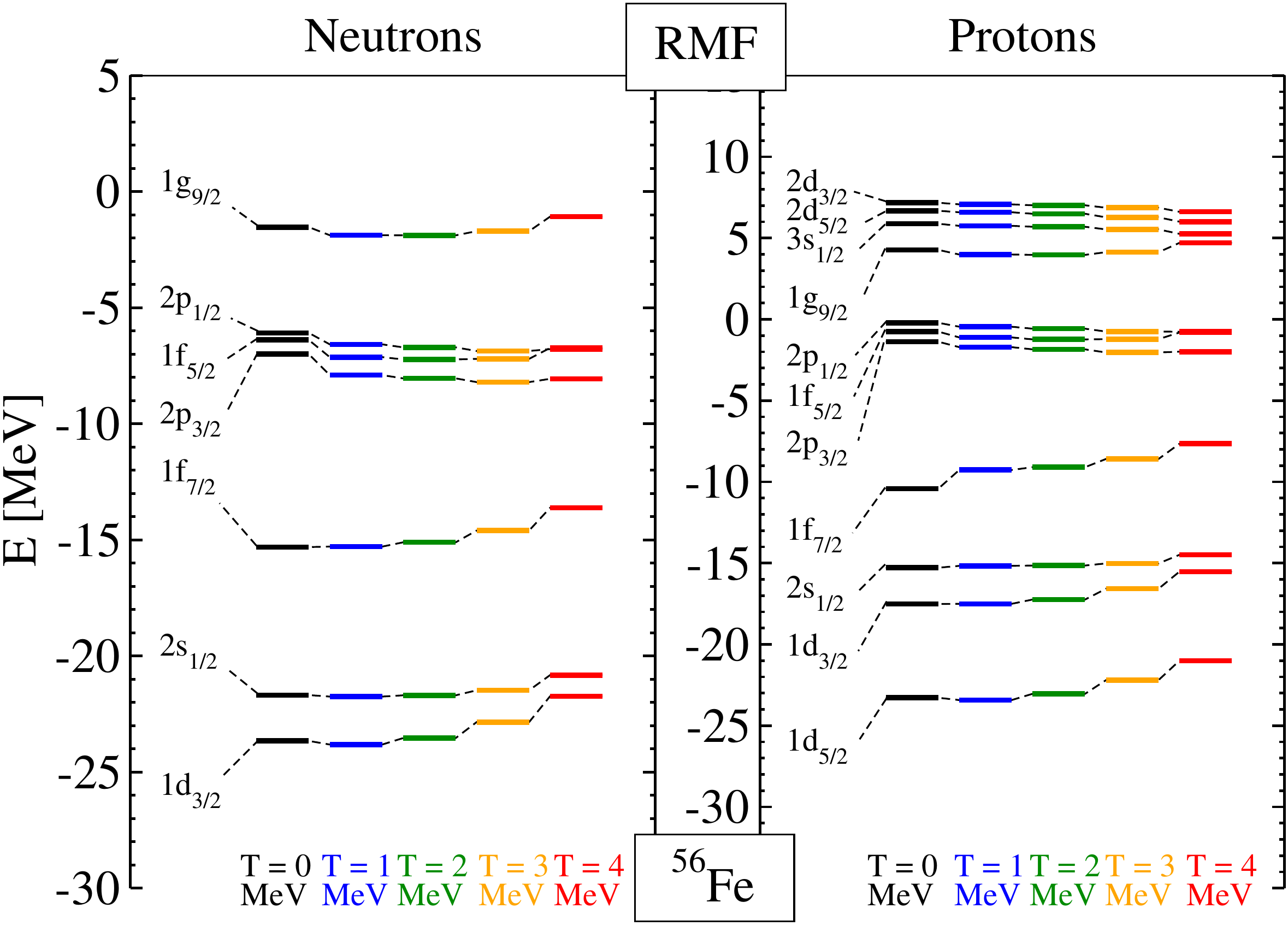}
\end{center}
\caption{The single-particle states in $^{56}$Fe at zero and finite temperatures calculated in the RMF approximation.}
\label{fe-RMF}%
\end{figure}
\begin{figure}
\begin{center}
\includegraphics[scale=0.35]{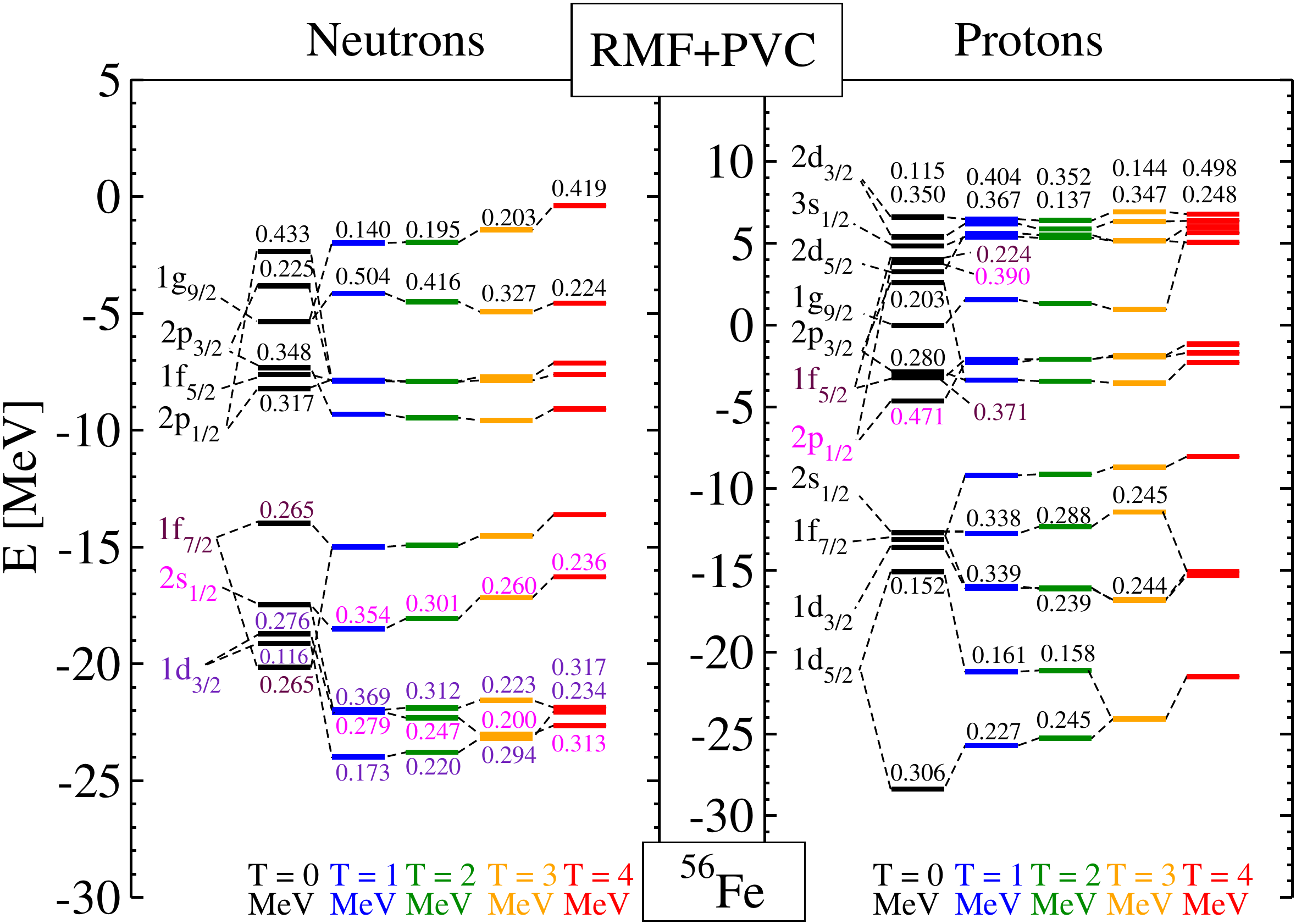}
\end{center}
\caption{The dominant fragments of the single-particle states in $^{56}$Fe at zero and finite temperatures calculated in the RMF+PVC approximation.}
\label{fe-RMF-PVC}%
\end{figure}

An example of a nucleus with strong pairing correlations in both neutron and proton subsystems is represented by $^{56}$Fe. Indeed, the first excited state in this nucleus is the 2$^+$ state at 846.78 keV followed by the 4$^+$ state at 2085.1 keV, according to the experimental measurements  \cite{nndc}. Although the relativistic quasiparticle RPA (RQRPA) can not reproduce accurately the excitation spectrum, it nevertheless returns the energies $E(2^+_1)$ = 1.12 MeV and $E(4^+_1)$ = 4.07 MeV, which stipulate strong PVC effects. As a result of the RMF+PVC calculations with pairing correlations for this nucleus at $T = 0$, we obtain strong fragmentation of the single-particle states even around the Fermi surface. This is reflected in Fig. \ref{fe-RMF-PVC}, where one can see quite a number of such states represented by pairs of their competing major fragments. A significant compression of both neutron and proton single-particle spectra relative to the RMF calculations shown in Fig. \ref{fe-RMF} is another consequence of the strong PVC in the broad energy region around the Fermi energy. As it can be seen in Fig. \ref{fe-RMF-PVC}, the first step of the temperature evolution of $^{56}$Fe is the phase transition from the superfluid to the non-superfluid state, which is indicated by the drastic decrease of the density of the single-particle spectra in both neutron and proton subsystems, when the temperature raises from $T = 0$ to $T = 1$ MeV. Further temperature evolution consists of a rather smooth redistribution of the spectroscopic strength between the major fragments.

\begin{figure}
\begin{center}
\includegraphics[scale=0.35]{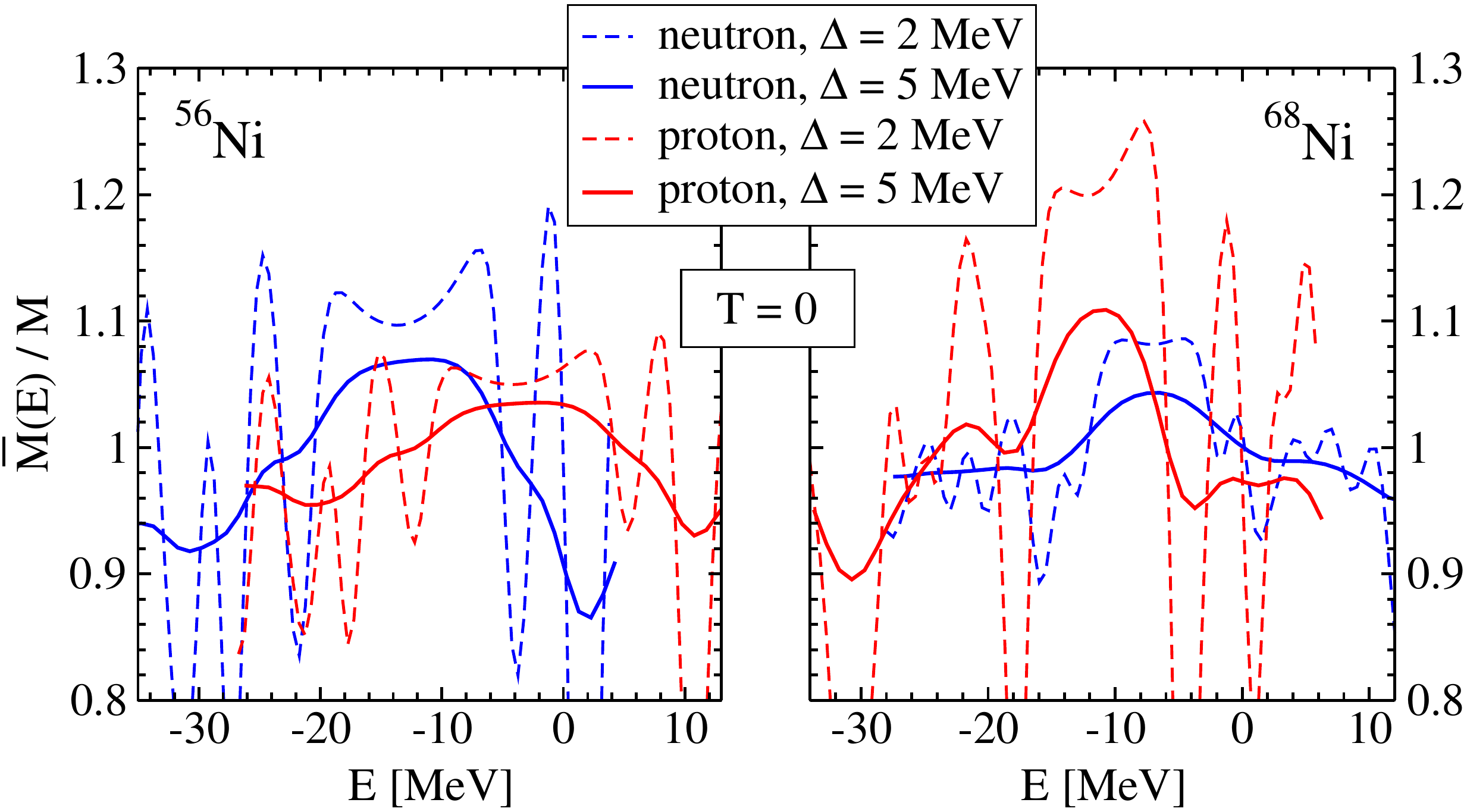}
\end{center}
\caption{Neutron (blue) and proton (red) dynamical effective masses in $^{56}$Ni (left) and $^{68}$Ni (right) computed with the imaginary parts of the energy variable $\Delta$ = 2 MeV (dashed curves)  and $\Delta$ = 5 MeV (solid curves) at $T = 0$. The dynamical effective masses are averaged over the single-particle states within 40 MeV energy window around the neutron and proton Fermi surfaces, respectively.}
\label{ni_Mef_D}%
\end{figure}

It is difficult, however, to assess the global evolution of the shell structure by looking only at the major fragments.
A very important characteristic of the strongly-coupled fermionic systems, namely the effective mass, is linked to both the single-particle level density and the nucleonic self-energy and, thus, can help evaluate the general trends. For relativistic systems, the authors of Ref. \cite{Jaminon1989} introduced the non-relativistic type effective mass, whose energy dependence 
at low energies is similar to that of non-relativistic systems \cite{Mahaux1981,Bortignon1982}:
\be
\frac{{\bar M}_{(k)}(\varepsilon)}{M} = 1 - \frac{d}{d\varepsilon}{\text Re}\Sigma^e_{(kk)}(\varepsilon),
\label{Mef}
\ee
where the energy argument is a complex number and $M$ is the mass of the bare nucleon. We can call the quantity ${\bar M}_{(k)}(\varepsilon)/{M}$ dynamical, or energy-dependent, effective mass. The indices in the brackets indicate the reduced matrix elements: $k = \{(k), m_k\}$, where $m_k$ is the projection of the total angular momentum on the quantization axis, which is commonly called magnetic quantum number in spherical symmetry. The effective mass averaged over the single-particle levels reads \cite{Bortignon1982}:
\be
\langle\frac{{\bar M}(\varepsilon)}{M}\rangle  = \sum\limits_{(k)}(2j_{(k)} + 1)\frac{{\bar M}_{(k)}(\varepsilon)}{M} / \sum\limits_{(k)}(2j_{(k)} + 1).
\ee  
In order to evaluate the nucleonic effective mass in nuclei, a finite imaginary part of the energy variable is used in Eq.  (\ref{Mef}): $\varepsilon = E + i\Delta$. The role of the imaginary part is to soften the singularities of the self-energy by averaging over the discrete single-particle spectrum. In this way, the self-energy acquires both real and significant imaginary parts. While the real part contributes to the effective mass as in Eq. (\ref{Mef}), the imaginary part plays the role of the optical potential.  The value of $\Delta$ should be thus associated with the average distance between the single-particle states. In our approach, for the medium-mass nuclei in the iron and nickel mass region the value of $\Delta$ = 5 MeV  satisfies this criterion. However, in order to have an idea about the sensitivity of the effective mass to this parameter, we have calculated it with  $\Delta$ = 2 MeV and $\Delta$ = 5 MeV for two isotopes of nickel, $^{56}$Ni and $^{68}$Ni, at T=0, where we have neglected the pairing correlations in the latter nucleus. The calculations and averaging were performed within symmetric 40 MeV intervals around the Fermi energies for both neutrons and protons. 
The results for the neutron and proton effective masses are displayed in Fig. \ref{ni_Mef_D}.
One can see that, indeed, the effective mass as a function of energy is sensitive to the averaging parameter, which reflects the non-continuity of the single-particle spectrum. Independently of that, the neutron and proton effective masses can be different, although both of them exhibit broad peaks around the corresponding Fermi energies, when a sufficiently large averaging is used. In the $N=Z$ nucleus $^{56}$Ni the neutron and proton effective masses show very similar behavior, although the neutron effective mass varies in a slightly larger range, than the proton one.  The situation is different in the neutron-rich nucleus $^{68}$Ni, where the proton effective mass demonstrates a remarkably stronger variation between its central and peripheral values. This result is consistent with the general trend, according to which  the effective mass increases with the density of states: indeed, one can conclude from Figs.  \ref{RMF} and \ref{RMF-PVC} that the overall density of the proton single-particle states in $^{68}$Ni is more affected by the PVC effects than the density of states in the neutron subsystem.    

\begin{figure}
\begin{center}
\includegraphics[scale=0.35]{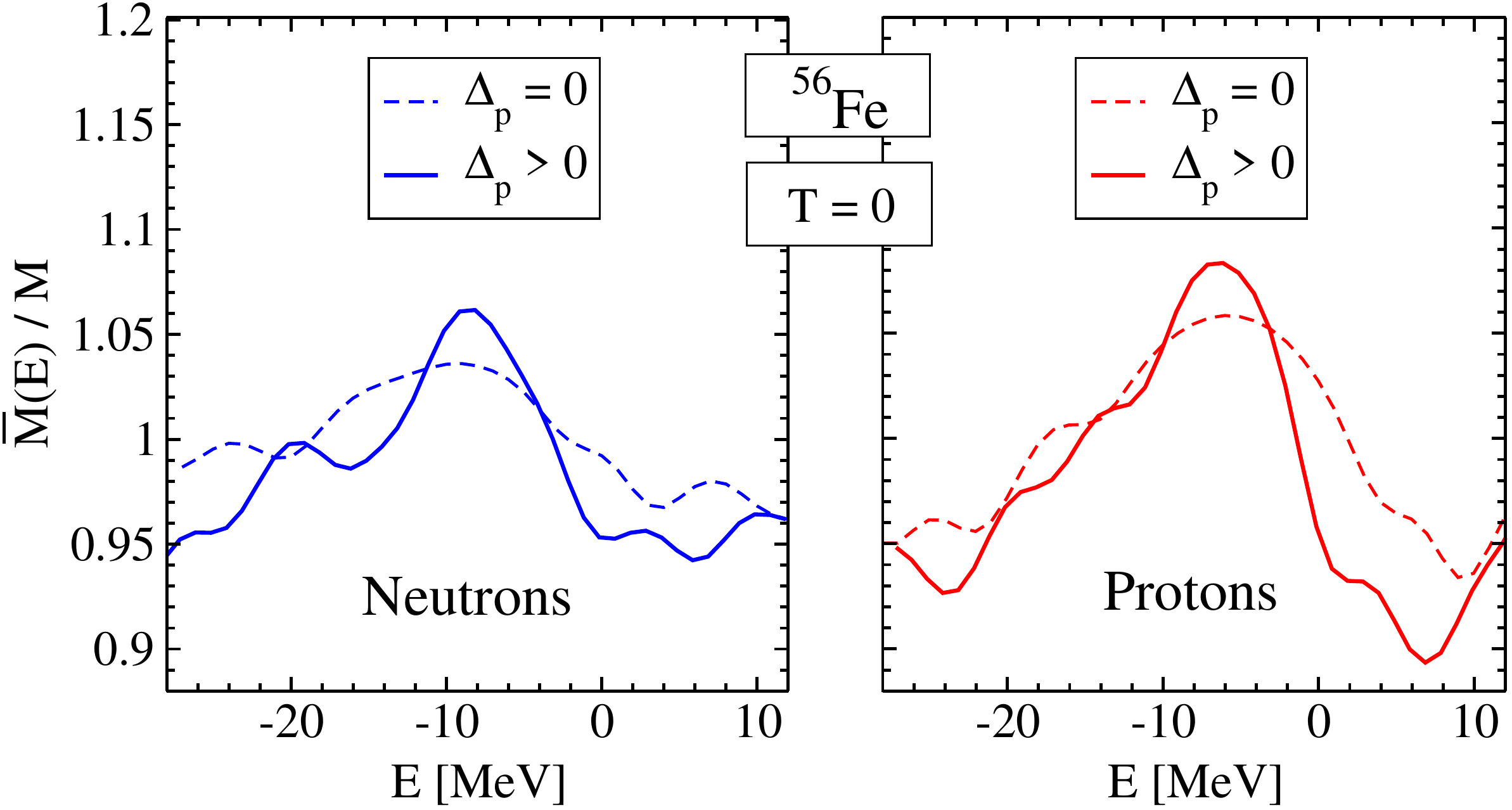}
\end{center}
\caption{The dynamical effective masses of neutrons and protons in $^{56}$Fe calculated with (solid curves) and without (dashed curves) pairing correlations at zero temperature. The value $\Delta$ = 5 MeV was used for the imaginary part of the energy variable. The symbol $\Delta_p$ stands for the superfluid pairing gap.}
\label{fe_Mef_p}%
\end{figure}

Fig. \ref{fe_Mef_p} illustrates the sensitivity of the dynamical effective mass to pairing correlations by displaying the neutron and proton effective masses in $^{56}$Fe at $T = 0$. As the neutron excess in this nucleus is small, the effective masses of neutrons and protons are represented by similar functions of energy, although the variation of the proton effective mass is somewhat stronger. Both effective masses show peaks in broad energy intervals around the Fermi surfaces. The widths of the peaks diminish when the pairing correlations are taken into account. The pairing correlations also cause stronger variations of the effective masses: the higher peak values and the lower values in the peripheral areas.  Overall, this result is consistent with the well-established proportionality of the dynamical effective mass to the density of states, which is higher around the Fermi surface, when pairing is included.

\begin{figure}
\begin{center}
\vspace{1cm}
\includegraphics[scale=0.30]{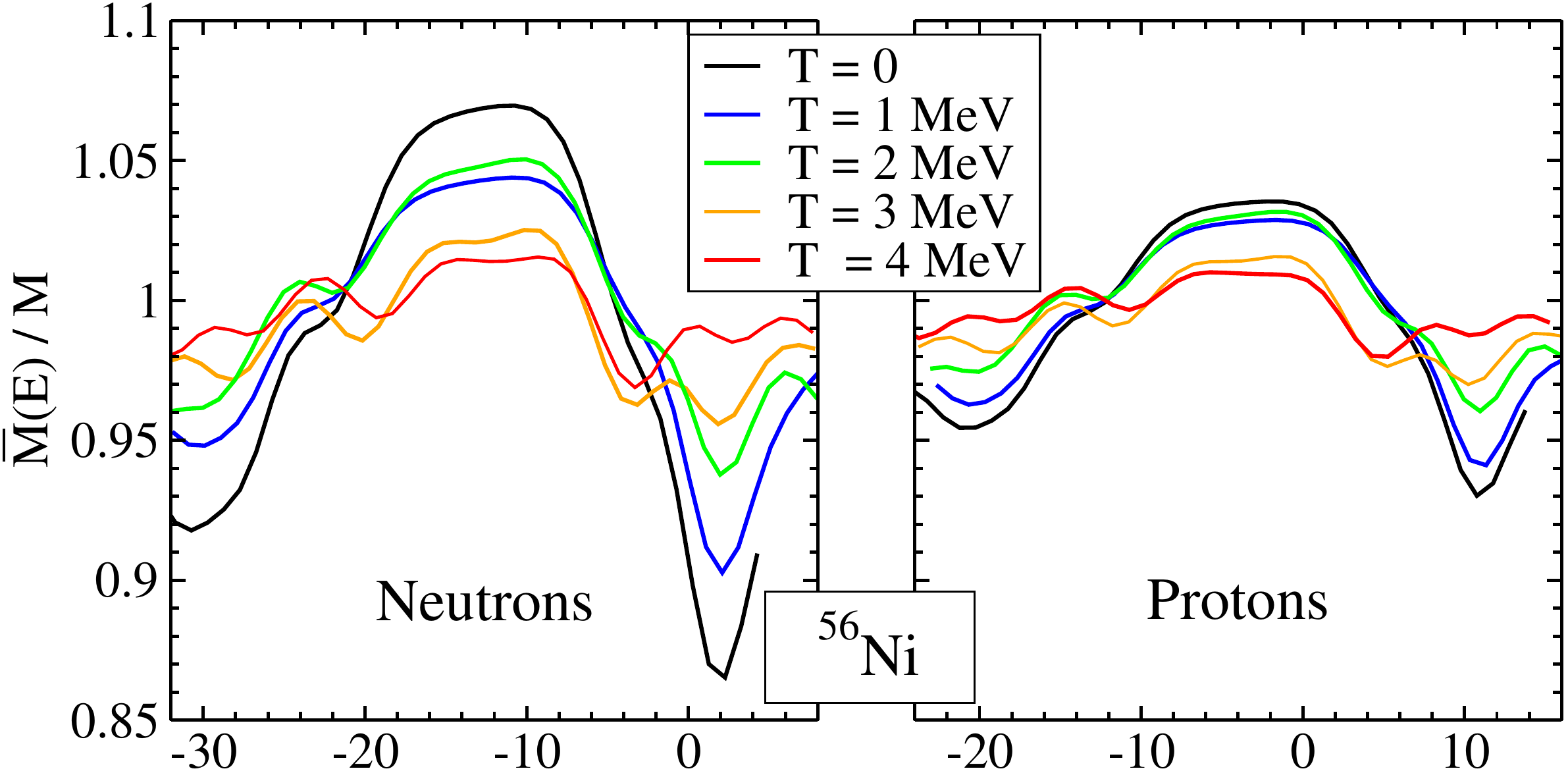}
\includegraphics[scale=0.30]{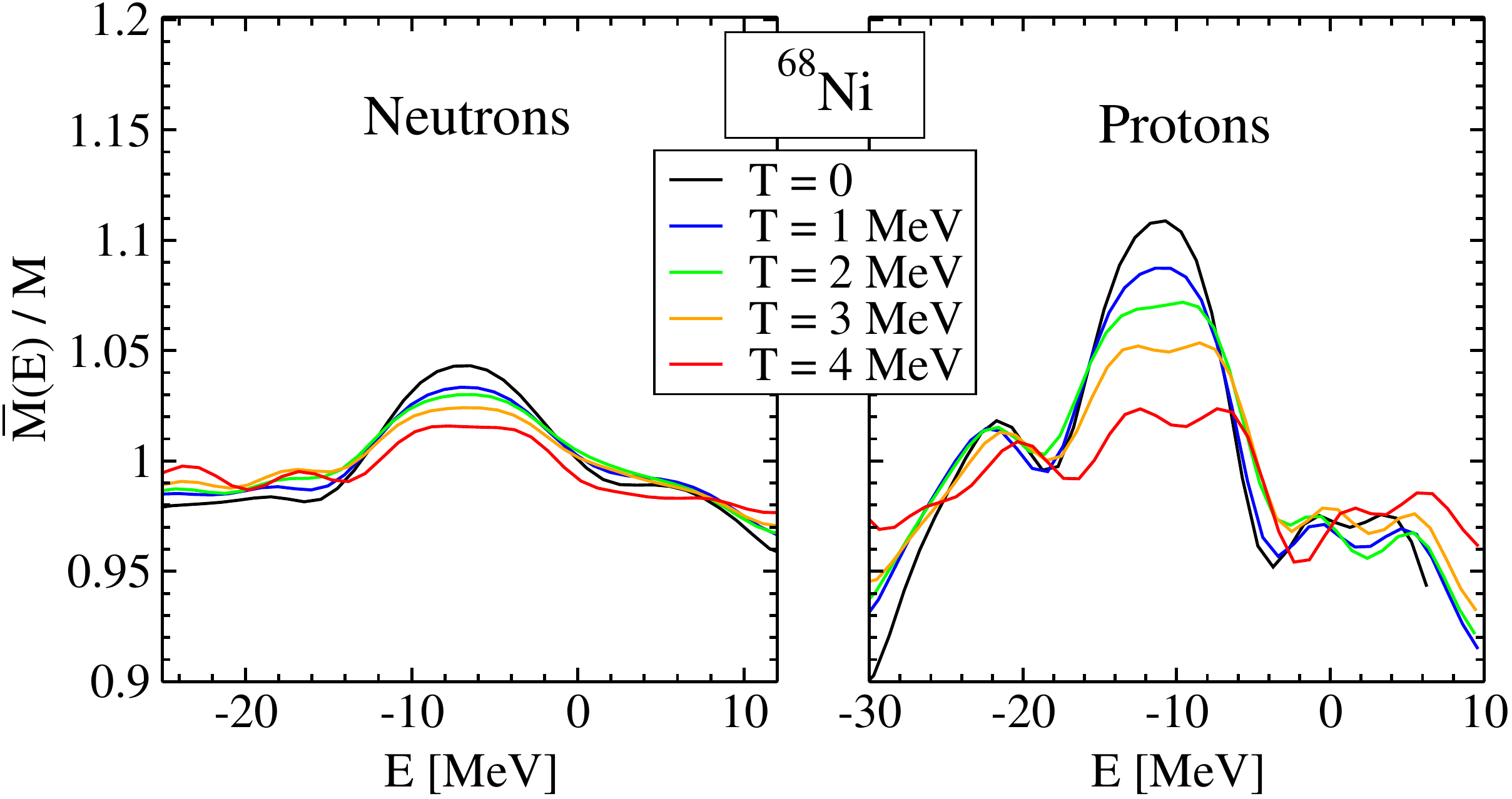}
\end{center}
\caption{Temperature evolution of the nucleon dynamical effective masses in $^{56,68}$Ni calculated with $\Delta$ = 5 MeV. 
}
\label{ni_Mef}%
\end{figure}

Finally, Fig. \ref{ni_Mef} illustrates the temperature evolution of the dynamical effective masses in $^{56,68}$Ni. As it was mentioned above, it is difficult to make definite conclusions on the evolution of the overall density of states by looking at the fragmented states themselves. The calculations of the effective masses, however, offer this opportunity. The results of calculations at various temperatures show a smooth evolution of the effective masses toward a more uniform energy dependence. With the temperature  increase, the effective masses in the peak region become less pronounced, while in the peripheral areas their values grow.     
In all the cases, when the temperature grows to higher values $T \geq$ 3 MeV, the dynamical effective masses tend to approach unity showing some minor oscillations around this value. The temperature evolution of the dynamical effective masses in $^{56}$Fe shows similar trends, that are consistent with the study of Ref. \cite{Donati1994}, although we have obtained overall smaller effective mass values at the Fermi surfaces.   

\section{Summary and outlook}
\label{Summary}

We have developed a many-body approach to describe fragmentation of the single-particle states in strongly-coupled fermionic systems at finite temperature. 
The dynamical finite-temperature self-energy is detailed for applications to atomic nuclei, where the leading fragmentation mechanism is the coupling to correlated particle-hole pairs, which represent emergent phonons of predominantly vibrational character.  
The Dyson equation with the particle-vibration self-energy has been solved numerically in the basis of the relativistic mean field for medium-mass nuclei, such as $^{56,68}$Ni and $^{56}$Fe, which play important roles in understanding stellar evolution. The temperature evolution of the fragmentation mechanism has been analyzed in detail by extracting the complete fragmented single-particle spectra in the 40 MeV window around the Fermi energies of the considered nuclei. Various scenarios realized for different types of states have been discussed.

To characterize the spectra globally, we have  computed the averaged dynamical neutron and proton effective masses as functions of energy at various temperatures. In cold nuclear systems, i.e. at zero temperature, the  dynamical effective masses show a clear bell-shaped behavior with maxima around the Fermi energy. We found that, depending on the interpretation and on the value of the averaging parameter, the variation of the dynamical effective mass can reach 10-20\% between the central and peripheral values, with respect to unity. Pairing correlations were found to sharpen the dynamical effective mass as a function of energy, while the temperature increase reduces the variation between the central and peripheral values.  

The obtained results may have some significance for astrophysical modeling of various stages of stellar evolution. As the nucleon effective mass is directly related to the density of states, entropy and symmetry energy, its correct temperature dependence can be important, in particular, for the core-collapse supernova modeling \cite{Donati1994}. The existing strategies to accommodate this dependence suggest simple parametrizations of the dynamical effective mass, however, some sensitivity studies would be helpful to establish whether its non-smooth oscillating behavior can be neglected on the global scale.  

\ 
 
\section{Acknowledgements}
This work is supported by the US-NSF Career Grant PHY-1654379.
%

\bibliography{Bibliography_Jun2020}
\end{document}